\makeatletter
\newcommand{\dontusepackage}[2][]{%
  \@namedef{ver@#2.sty}{9999/12/31}%
  \@namedef{opt@#2.sty}{#1}}
\makeatother
\dontusepackage{subfigure}

\documentclass[]{sigchi}

\usepackage{lmodern}
\usepackage{amssymb,amsmath}
\usepackage{ifxetex,ifluatex}
\usepackage[usenames,dvipsnames]{color}
\usepackage{fixltx2e} 
\ifnum 0\ifxetex 1\fi\ifluatex 1\fi=0 
  \usepackage[utf8]{inputenc}
\else 
  \ifxetex
    \usepackage{mathspec}
    \usepackage{xltxtra,xunicode}
  \else
    \usepackage{fontspec}
  \fi
  \defaultfontfeatures{Mapping=tex-text,Scale=MatchLowercase}
  
\fi
\IfFileExists{upquote.sty}{\usepackage{upquote}}{}
\IfFileExists{microtype.sty}{%
\usepackage{microtype}
\UseMicrotypeSet[protrusion]{basicmath} 
}{}
\usepackage{listings}
\usepackage{longtable,booktabs}
\usepackage{graphicx}
\makeatletter
\def\maxwidth{\ifdim\Gin@nat@width>\linewidth\linewidth\else\Gin@nat@width\fi}
\def\maxheight{\ifdim\Gin@nat@height>\textheight\textheight\else\Gin@nat@height\fi}
\makeatother
\setkeys{Gin}{width=\maxwidth,height=\maxheight,keepaspectratio}
\usepackage[font={small,bf},skip=2pt]{caption}
\usepackage{float}
\usepackage{subfig}
\def\plainauthor{Jérémy Frey, Maxime Daniel, Julien Castet, Martin Hachet, Fabien Lotte}

\ifxetex
  \usepackage[setpagesize=false, 
              unicode=false, 
              xetex]{hyperref}
\else
  \usepackage[unicode=true]{hyperref}
\fi
\hypersetup{breaklinks=true,
            bookmarks=true,
            pdfauthor={\plainauthor},
            pdftitle={Framework for Electroencephalography-based Evaluation of User Experience},
            colorlinks=true,
            citecolor=black,
            urlcolor=blue,
            linkcolor=black,
            pdfborder={0 0 0}}
\usepackage{balance}

\title{Framework for Electroencephalography-based Evaluation of User Experience}

\numberofauthors{5}
\author{
  \alignauthor Jérémy Frey\\
    \affaddr{Univ. Bordeaux, France}\\
    \email{jeremy.frey@inria.fr}\\
  \alignauthor Maxime Daniel\\
    \affaddr{Immersion SAS, France}\\
    \email{maxime.daniel@u-bordeaux.fr}\\ 
  \alignauthor Julien Castet\\
    \affaddr{Immersion SAS, France}\\
    \email{julien.castet@immersion.fr}\\ 
  \alignauthor Martin Hachet\\
    \affaddr{Inria, France}\\
    \email{martin.hachet@inria.fr}\\
  \alignauthor Fabien Lotte\\
    \affaddr{Inria, France}\\
    \email{fabien.lotte@inria.fr}\\
}

\date{}

\usepackage{times}

\usepackage{suffix}

\usepackage{subfig}

 \toappear{Publication rights licensed to ACM. ACM acknowledges that this contribution was authored or co-authored by an employee, contractor or affiliate of a national government. As such, the Government retains a nonexclusive, royalty-free right to publish or reproduce this article, or to allow others to do so, for Government purposes only \\
 \emph{CHI'16}, May 07 - 12, 2016, San Jose, CA, USA \\
Copyright is held by the owner/author(s). Publication rights licensed to ACM. \\
ACM 978-1-4503-3362-7/16/05...\$15.00 \\
DOI: http://dx.doi.org/10.1145/2858036.2858525}

\begin{document}

      \teaser{
      \includegraphics[width=\textwidth]{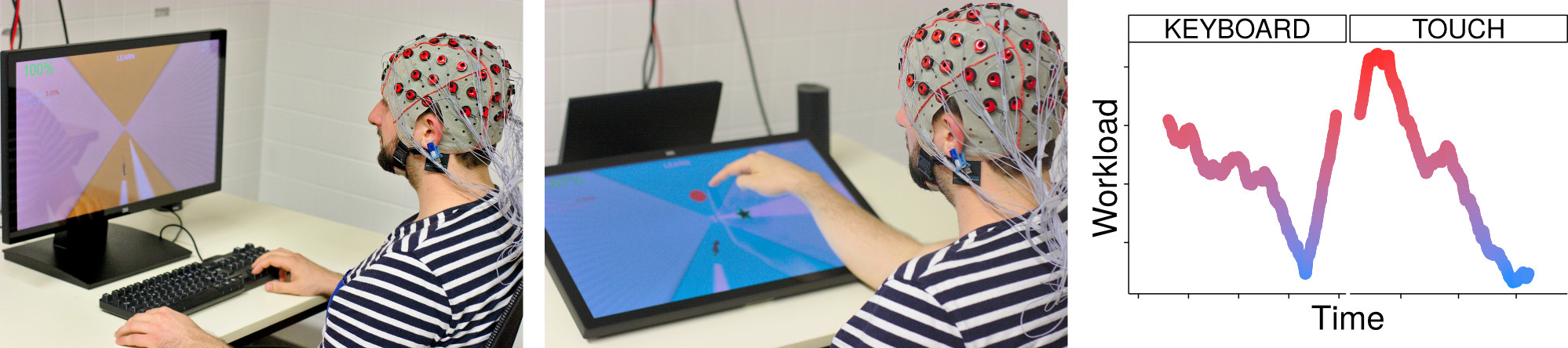}
      \caption{We demonstrate how electroencephalography can be used to evaluate
human-computer interaction. For example, a keyboard (\emph{left}) can be
compared with a touch interface (\emph{middle}) using a continuous
measure of mental workload (\emph{right}, here participant 4).\label{fig:teaser}}
    }
  
\maketitle
\begin{abstract}
Measuring brain activity with electroencephalography (EEG) is mature
enough to assess mental states. Combined with existing methods, such
tool can be used to strengthen the understanding of user experience. We
contribute a set of methods to estimate continuously the user's mental
workload, attention and recognition of interaction errors during
different interaction tasks. We validate these measures on a controlled
virtual environment and show how they can be used to compare different
interaction techniques or devices, by comparing here a keyboard and a
touch-based interface. Thanks to such a framework, EEG becomes a
promising method to improve the overall usability of complex computer
systems.
\end{abstract}

\keywords{
      EEG;
      HCI Evaluation;
      Workload;
      Attention;
      Interaction errors;
      Neuroergonomy}

      \category{H.5.2}{User Interfaces}{Evaluation/methodology}

\def \citep {\protect\cite}



\makeatletter
\def\url@leostyle{%
  \@ifundefined{selectfont}{\def\UrlFont{\sf}}{\def\UrlFont{\small\bf\ttfamily}}}
\makeatother
\urlstyle{leo}

\def\pprw{8.5in}
\def\pprh{11in}

\setlength{\paperwidth}{\pprw}
\setlength{\paperheight}{\pprh}
\setlength{\pdfpagewidth}{\pprw}
\setlength{\pdfpageheight}{\pprh}

\definecolor{linkColor}{RGB}{6,125,233}
\hypersetup{%
  bookmarksnumbered,
  colorlinks,
  citecolor=black,
  filecolor=black,
  linkcolor=black,
  urlcolor=linkColor,
  breaklinks=true,
}

\WithSuffix\newcommand\caption*{\caption}

\newcommand{\leveldown}
  {\let\section\subsection%
   \let\subsection\subsubsection%
   \let\subsubsection\paragraph%
   \let\paragraph\subparagraph%
  }
  
\newcommand{\levelup}
  {\let\subparagraph\paragraph%
   \let\paragraph\subsubsection%
   \let\subsubsection\subsection%
   \let\subsection\section%
  }

\let\oldsubfloat\subfloat
\renewcommand*{\subfloat}{\hfill\oldsubfloat}

\section{Introduction}\label{introduction}

In practice, a tool is only as good as one's ability to assess it. For
instance, evaluations in Human-Computer Interaction (HCI) usually rely
on inquiries -- e.g., questionnaires or think aloud protocols -- or on
users' behavior during the interaction-- e.g., reaction time or error
rate. However, while both types of methods have been used successfully
for decades, they suffer from some limitations. Inquiries are prone to
be contaminated by ambiguities \citep{Nisbett1977} or may be affected by
social pressure \citep{Picard1995}. It is also very difficult to gain
real-time insights without disrupting the interaction. Indeed, think
aloud protocol distracts users and questionnaires can be given only at
specific time points, usually at the end of a session -- which leads to
a bias due to participants' memory limitations \citep{Kivikangas2010}.
On the other hand, metrics inferred from behavioral measures can be
computed in real-time, but are mostly quantitative. They do not provide
much information about users' mental states. For example, a high
reaction time can be caused either by a low concentration level or by a
difficult task \citep{Berka2007, Hart1988}.

Recently, is has been suggested that portable brain imaging techniques
-- such as electroencephalography (EEG) and functional near infrared
spectroscopy (fNIRS) -- have the potential to address these limitations
\citep{Fairclough2009a, Pikea2012, Frey2014a}. fNIRS has been studied to
assess users' workload, for example to evaluate user interfaces
\citep{Hirshfield2009a} or different data visualizations
\citep{Peck2013}. However, most of these works are evaluating passive
tasks or very basic interactions that are not ecological, especially as
user interfaces and interactions are getting more complex.

In this paper, not only do we detail a framework for HCI designers to
gain further insights about user experience using brain signals -- here
from EEG -- but we also validate this framework on an actual and
realistic task. We show that they can be used to compare different HCI,
in an environment that provides many simulations and where participants
are engaged in rich interactions.

In the following sections, we describe the virtual environment that we
developed, specifically aimed at validating the use of EEG as an
evaluation method for HCI. We validated the workload induced by our
environment in a first study, with NASA-TLX questionnaires
\citep{Hart1988}. Then, we detail how EEG can be used to assess 3
different constructs: workload, attention and error recognition.
Finally, during the main study we employ EEG recordings to measure
continuously such workload, altogether with the attention level of
participants toward external stimuli and the number of interaction
errors they perceived, while they interact with our virtual environment.
We took the opportunity of these EEG-based measures to compare a
keyboard input to a touch screen input.

To summarize, our main contributions are:

\begin{enumerate}
\def\labelenumi{\arabic{enumi}.}
\itemsep1pt\parskip0pt\parsep0pt
\item
  To validate the use of EEG as a continuous HCI evaluation method
  thanks to controlled and realistic interaction tasks we designed.
\item
  To demonstrate how such tool can assess which of the tested
  interaction technique is better suited for a particular environment.
\item
  To propose a framework that can be easily replicated to improve
  existing interfaces with little or no modifications.
\end{enumerate}

\subsection{Related work}\label{related-work}

Since a few years, brain imaging has been used in HCI to deepen the
understanding of users, thanks notably to the spread of affordable and
lightweight devices. For instance, EEG and fNIRS are particularly well
suited for mobile brain imaging \citep{Mehta2013, Cutrell2008}. These
techniques can be employed to study the spatial focus of attention
\citep{Trachel2013} or to identify system errors
\citep{Vi2012, putze2015design}.

EEG and fNIRS are opportunities to assess the overall usability of a
system and improve the ergonomics of HCI. In a recent work we showed
preliminary results regarding the evaluation of mental workload during a
3D manipulation task \citep{Wobrock2015}. After being processed, EEG
signals highlighted which parts of the interaction induced a higher
mental workload. In the present paper we go much beyond, exploring a
\emph{continuous} index of workload as well as two others constructs,
namely attention and error recognition. We also rigorously validate such
indexes, and study them in light of behavioral measures (performances
and reaction times) and inquiries (NASA TLX questionnaire).

Attention refers to the ability to focus cognitive resources on a
particular stimulus \citep{Kivikangas2010}. In HCI, measuring the
attention level could help to estimate how much information users
perceive. In the present work the measure of attention relates to
inattentional blindness; i.e.~it concerns participants' capacity to
process stimuli irrelevant to the task \citep{Cartwright-Finch2007}.

Error recognition relates to the detection by users of an outcome
different from what is expected \citep{Nieuwenhuis2001}. We focused on
\emph{interaction} errors \citep{Ferrez2008}, which arise when a system
reacts in an unexpected way, for example if a touch gesture is badly
recognized. Interaction errors enable to assess how intuitive a UI is,
and they are hardly measurable by another physiological signal than EEG.
The combined measure of workload, attention and error recognition
constitutes a powerful complementary evaluation tool for people who
design new interaction techniques.

Even though commercial solutions, such as the B-Alert system\footnote{http://www.advancedbrainmonitoring.com/},
are already pointing this direction, they are validated in the
literature with lab tasks only \citep{Berka2007}. Our work, on the other
hand, is closer to the field. More importantly, as opposed to
proprietary software, our methodology is transparent and our
multidimensional index of user experience can be easily replicated.

\section{Virtual 3D maze}\label{virtual-3d-maze}

The 3D virtual environment that we built uses gamification
\citep{Deterding2011} to increase users' engagement and ensure better
physiological recordings \citep{Flatla2011}. Such a virtual environment
also enables us to assess workload, attention and error recognition
during ecological and realistic interaction tasks. Indeed, such
constructs are traditionally evaluated during controlled lab experiments
based on protocols from psychology that are vastly different from an
actual interaction task, see, e.g., \citep{Grimes2008}.

\begin{figure*}
\centering
\subfloat[\label{fig:maze-remind}]{\includegraphics[width=0.320\hsize]{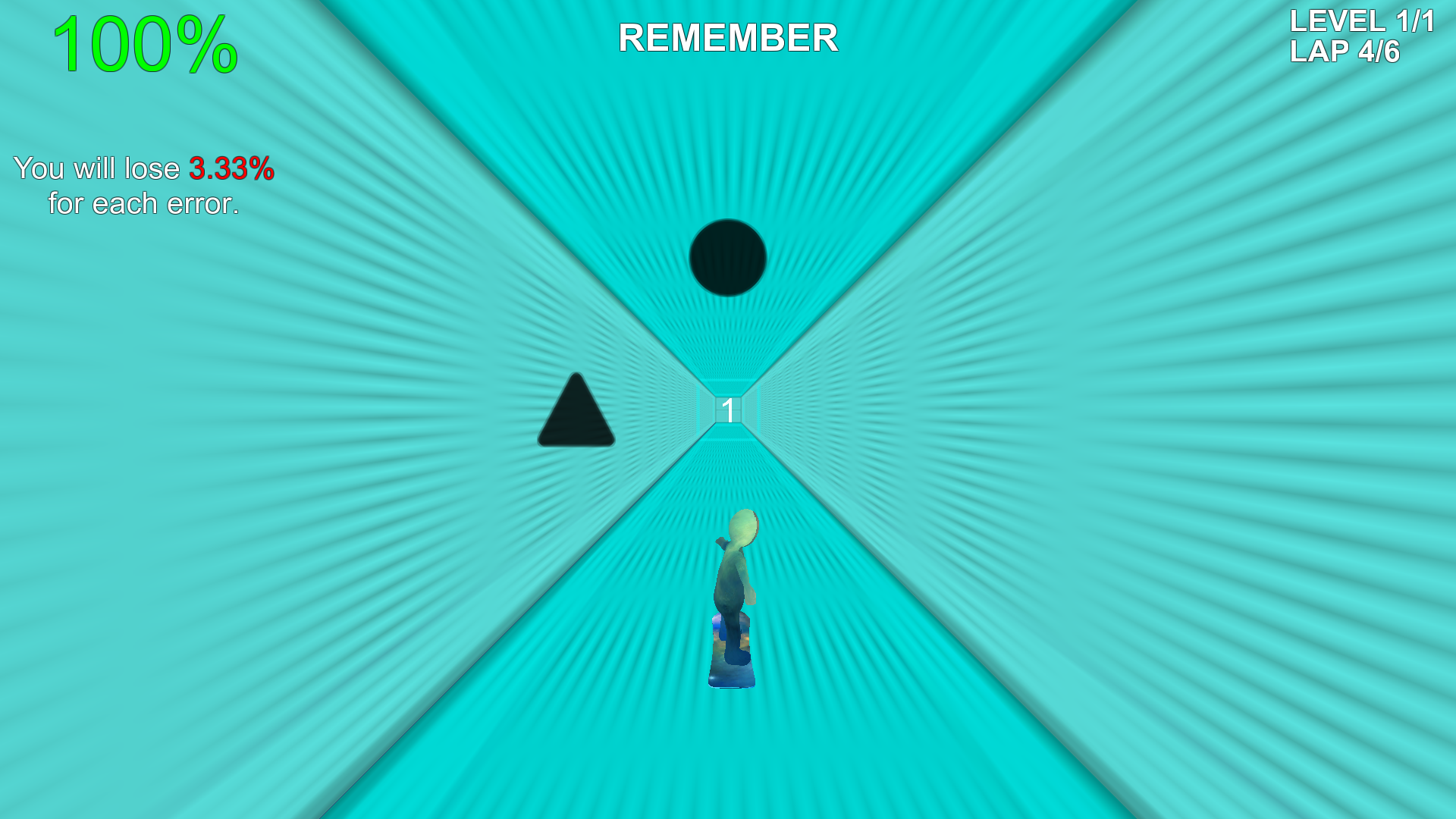}}
\subfloat[\label{fig:maze-hint}]{\includegraphics[width=0.320\hsize]{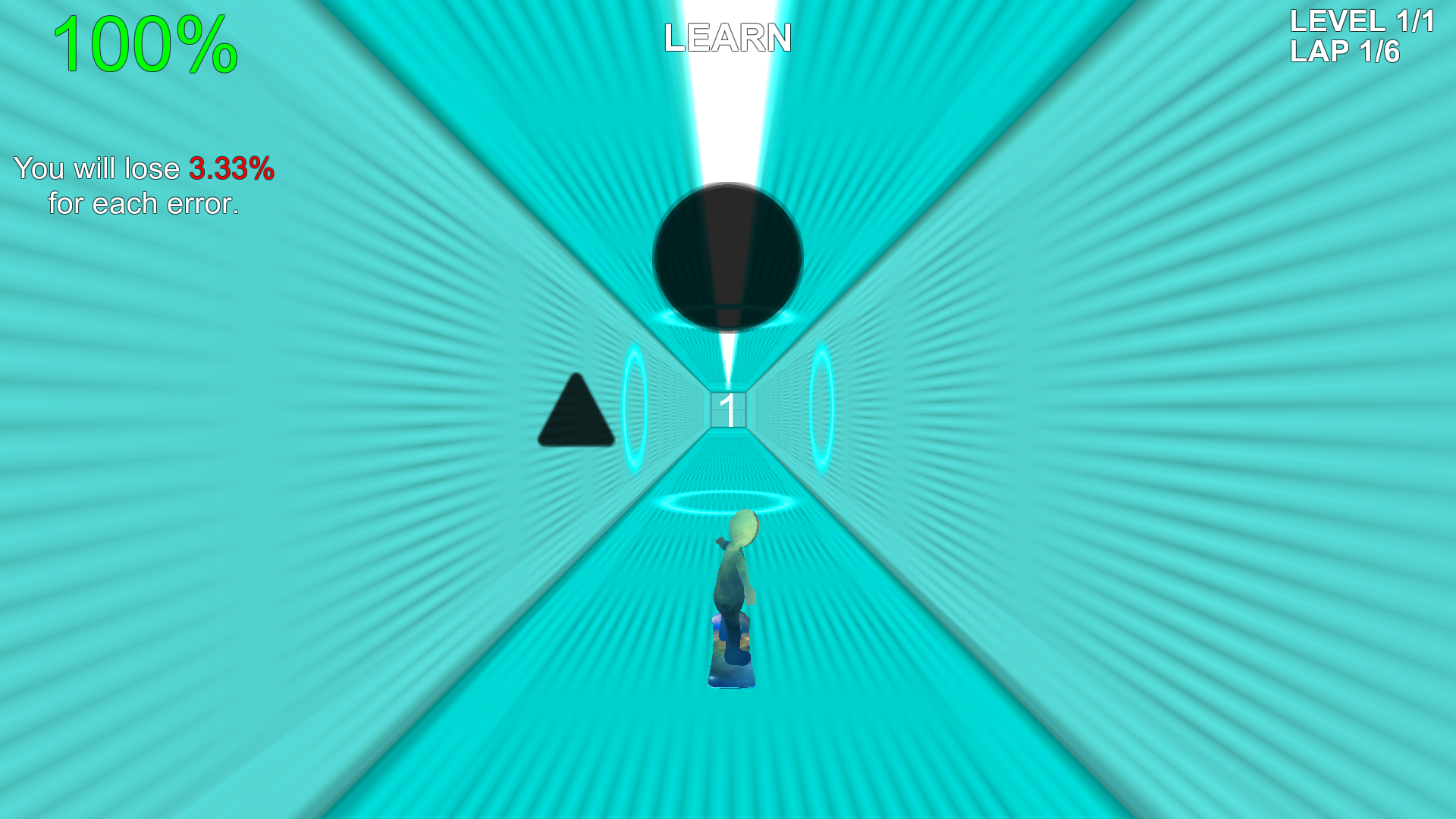}}
\subfloat[\label{fig:maze-orientation}]{\includegraphics[width=0.320\hsize]{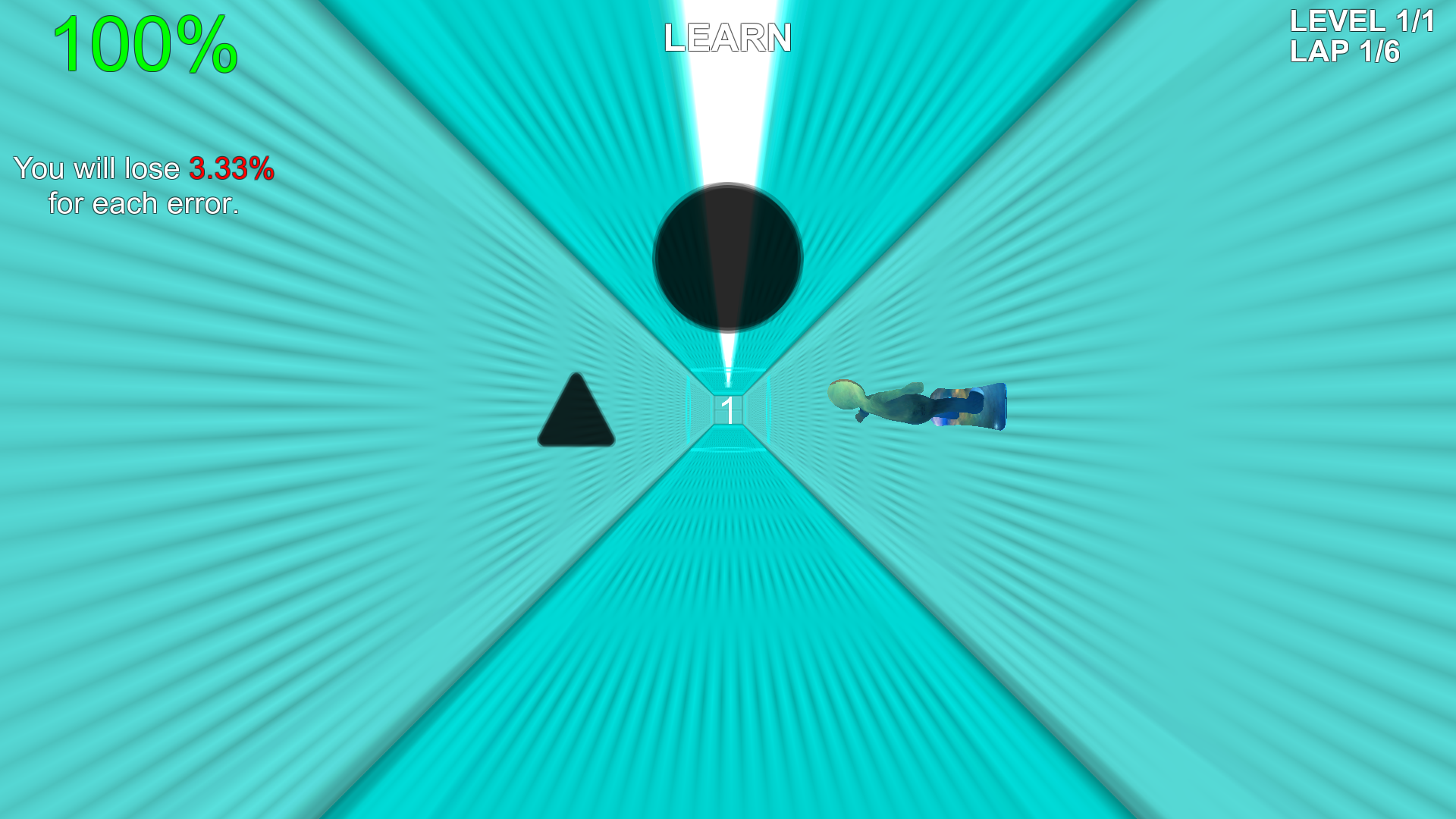}}
\caption{The virtual environment, where players control a character that
moves by itself inside a 3D maze. \emph{A}: Symbols appear in each
tunnel to indicate the possible directions for the next turn; players
have to select a particular sequence of symbols/directions. \emph{B}:
During the ``learning'' phase, the correct direction is highlighted by a
breadcrumb trail and the associated symbol bounces (here the disc on
top). \emph{C}: Controls depend on the position of the character. If the
character is on the right side, players have to press \emph{right} in
order to go \emph{up}.}\label{fig:maze}
\end{figure*}

\subsection{Overall description}\label{overall-description}

The virtual environment takes the form a maze where players have to
learn and reproduce a path by triggering directions at regular intervals
(see Figure~\ref{fig:maze}). A character displayed with a third person
perspective moves by itself at a predefined speed inside orthogonal
tunnels. Soon after the character enters a new tunnel, symbols appear
on-screen. Those symbols are basic 2D shapes, such as squares, circles,
triangles, diamonds or stars, and their positions (bottom, top, left or
right) indicate which directions are ``opened''. Players must select one
of these symbols before the character reaches the end of an
intersection, either by pressing a key or touching the screen. If users
respond too early, i.e., before symbols appeared, too late, or if they
select a direction that does not exist, they loose points and the
character ``dies'' by smashing against a wall, respawning soon after at
the beginning of the current tunnel.

The main element of the gameplay consists in selecting the directions in
the correct order. Indeed, one level comprised two phases. During the
``learning'' phase a particular sequence of symbols is highlighted; at
each symbols' appearance one of them is bouncing to indicate the correct
direction. Another cue takes the form of a ``breadcrumb trail'', a beam
of light that precedes the character and points to the right direction
(see Figure~\ref{fig:maze-hint}). Selecting an available but incorrect
direction does not result in the character's ``death'' but leads to a
loss of points. A visual feedback is given to users when they select a
direction: the corresponding symbol turns green if the choice is correct
and red otherwise. When the end of the maze is reached, the character
loops over the entire path so that players have another opportunity to
learn the sequence. When the training phase ends, the ``recall'' phase
follows. The symbols are identical but the cues are no more displayed;
players have to remember by themselves the right path. Symbols position
in each tunnel and symbols sequence are randomly drawn when a new level
starts.

Beside learning a sequence, the principal challenge comes from
\emph{how} the directions are selected. The third-person view fulfills a
purpose: the input device that users are controlling -- i.e.~keyboard or
touch screen -- is mapped to the \emph{character position}. Since the
character is a futuristic surfer that defies the law of gravity, it
slides by itself from the bottom of the tunnel to one of the walls or to
the ceiling from time to time. In this latter situation, when the
character is upside down, commands are inverted compared to what players
are used to, even though symbols remain in the same positions. This game
mechanism stresses spatial cognition abilities; users have to constantly
remain aware of two different frames of reference. For example, if
``up'' and ``left'' directions are open in a given tunnel and if the
character's position -- controlled by the application -- is on the right
wall, as illustrated in Figure~\ref{fig:maze-orientation}, users have to
press \emph{right} to go \emph{up}. This discrepancy between input and
output is a reminder of the problematic often observed with 3D user
interfaces, where most users manipulate a device with 2 degrees of
freedom (DOF), such as a mouse, to interact with a 6 DOF environment.

The combination of the game design and game mechanisms herein described
offers a wide variety of elements that we put in use so as to
investigate users' mental states. In particular, we detail below how we
tuned the game elements to manipulate the user's workload and attention
in controlled ways as well as to trigger interaction errors. Knowing
which constructs value (e.g.~high or low workload) to expect, we can
validate whether our EEG-based estimates during interaction match these
expectations, and thus whether they are reliable.

\subsection{Manipulating workload}\label{manipulating-workload}

Our virtual environment possesses several characteristics that could be
used to induce different levels of mental workload. We can notably
adjust 4 parameters:

\begin{itemize}
\itemsep1pt\parskip0pt\parsep0pt
\item
  \emph{Maze depth}: the number of tunnels players have to cross before
  reaching the end of the maze, hence the length of the symbols sequence
  they have to learn. More symbols to be held in the working memory
  increases workload \citep{Grimes2008, Sternberg1966}.
\item
  \emph{Number of directions}: at each intersection, up to 4 directions
  are ``opened'' in the maze; the complexity of the symbols sequence
  grows as this number increases.
\item
  \emph{Game speed}: the pace of the game can be adjusted to increase
  temporal pressure. When the speed increases symbols appear sooner and
  users must respond quicker, thus increasing overall stress
  \citep{Hart1988, Maule1997}. In the easiest level the character spends
  6s in a tunnel and players must respond within 3s after symbols
  appearance; in the hardest level a tunnel lasts 2s and players have 1s
  to choose a symbol.
\item
  \emph{Spatial orientation}: in order to keep selecting the correct
  directions, users have to perform a mental rotation if the character
  they control jumps from the floor to the walls or to the ceiling.
  Furthermore, they need to update their frame of reference as often as
  the character shifts from one side to another. Depending on the
  spatial ability of users, this mechanism can cause an important
  cognitive load \citep{Poor2013}.
\end{itemize}

We used those mechanisms and dimensions to create 4 different difficulty
levels for the game: ``EASY'', ``MEDIUM'', ``HARD'' and ``ULTRA'' (see
Table~\ref{tab:difficulties}). These levels affect mostly (symbolic)
memory load and time pressure. Indeed, the 3D maze is more about
remembering a sequence of symbols or directions rather than spatial
navigation per se. Because randomization could create loops in the maze
topography and since there were no landmarks, it is unlikely that
participants were able to adopt an allocentric strategy.

While the EASY level is designed to be completed with very little
effort, the ULTRA level, on the other hand, is designed to sustain a
very high level of workload, up to the point that it is barely possible
to complete it with no error. While during EASY levels there is no need
to perform mental rotations and players have to memorize only 2 symbols
that are constrained to either left and right directions, in ULTRA
levels the frame of reference changes between each selection and the
sequence reaches 5 symbols that could appear in all 4 directions,
\emph{and} players have to react thrice as fast. No matter the level,
players had 3 ``loops'' to learn the maze and another set of 3 loops to
reproduce the path.

\begin{table}
\centering
\begin{tabular}{lllll}
\toprule\addlinespace
Difficulty & Depth & Directions & Resp. time &
Orientation\tabularnewline
\midrule
EASY & 2 & 2 & 3s & 0\%\tabularnewline
MEDIUM & 4 & 3 & 2.5s & 30\%\tabularnewline
HARD & 5 & 4 & 2s & 60\%\tabularnewline
ULTRA & 5 & 4 & 1s & 100\%\tabularnewline
\bottomrule
\end{tabular}
\caption{Four difficulty levels are created by leveraging on game
mechanisms. \emph{Depth}: number of directions/symbols players have to
learn. \emph{Directions}: number of possible directions at each
intersection. \emph{Response time}: how much time players have to
respond after symbols appearance. \emph{Orientation}: percentage chance
that the controlled character changes its
orientation.}\label{tab:difficulties}
\end{table}

\subsection{Assessing attention}\label{assessing-attention}

We relied on stimuli not congruent to the main task in order to probe
for inattentional blindness, using the ``oddball'' paradigm. The oddball
paradigm is often employed with EEG as the appearance of rare (i.e.
``odd'') stimuli among a stream of frequent stimuli (i.e.~distractors)
triggers a particular event-related potential (ERP) within EEG signals
\citep{Coull1998}. ERP are ``peaks'' and ``valleys'' in EEG recordings,
and the amplitude of some of them decreases as users are less attentive
to stimuli.

Our protocol uses audio stimuli. It is based on \citep{Burns2015}, which
studied the immersion of video game players. In our virtual environment,
while users' characters were navigating in the maze, sounds were played
at regular intervals, serving as a background ``soundtrack'' that was
consistent with the user experience. 20\% of these sounds had a high
pitch (odd event) and the remaining 80\% had a low pitch (distracting
events) -- this proportion is on par with the literature
\citep{Burns2015, Ferrez2008}.

Our hypothesis is that the attention level of participants toward sounds
-- as measured with the oddball paradigm -- should decrease as the
workload increase, since most of their cognitive resources will be
allocated to the main task during the most demanding levels.

\subsection{Assessing error
recognition}\label{assessing-error-recognition}

EEG could be used to measure interaction errors, i.e.~errors originating
from an incorrect response of the user interface, that differs from what
users were expecting \citep{Ferrez2008}. Interaction errors are of
particular interest for HCI evaluation since they could account for how
intuitive an interface is \citep{Frey2014a}. In order to test the
feasibility of such measure, we decided to implement two different
interaction techniques. Both of them use discrete events --
i.e.~symbols' selection -- so that we could more easily synchronize EEG
recordings with in-game events later on.

The first technique uses \emph{indirect} interactions by the mean of a
keyboard (Figure \ref{fig:teaser}, left). In due time, left, right, up
or down arrow keys are used to send the character in the tunnel that is
situated to \emph{its} left, right, top or bottom. Indeed, we have seen
previously that in our virtual environment players have to orientate
themselves depending of the position of the character. If the character
is moving on the sides, players have to perform a mental rotation of
90°, if it is on the ceiling then the angle is 180°, i.e.~commands are
inverted.

The second technique uses \emph{direct} interaction by the mean of a
touch screen (Figure \ref{fig:teaser}, middle). Usually, with touch
screen, pointing is co-located with software events, since users can
directly indicate where they want to interact. However, in our case, we
decided to mimic exactly the behavior of the keyboard interface. That is
to say that with the touch screen as well players have to orientate
themselves depending on the position of the character. Hence, if the
character is positioned on the \emph{left}, players have to touch the
\emph{right} fringe of the screen in order to go \emph{up}. This is
mostly counter-intuitive since players have to inhibit the urge to point
to the actual direction they want to go; there is a cognitive
dissonance.

Since in our experimental design the use of the direct (touch-based)
interaction is counter-intuitive, we hypothesize that it will lead to an
overall higher number of interaction errors compared to the indirect
interface (keyboard).

\section{Pilot study: validation of the induced workload
level}\label{pilot-study-validation-of-the-induced-workload-level}

We designed our virtual environment as a test-bench aimed at inducing
several mental states within users, notably, different workload levels.
Thus, we had to formally validate the mental workload that each game
level seeks to induce. As such, we conducted a pilot study -- separate
from the main study to alleviate the protocol of the latter --, with no
physiological recordings but using the NASA-TLX questionnaire
\citep{Hart1988}, a well established questionnaire that accounts for
workload.

\subsection{Protocol}\label{protocol}

15 participants took part in this study (4 females), mean age 24.53 (SD:
3.00). We used a within-subject design; all participants answered for
all 4 difficulty levels. The gaming session used the keyboard and
started with 2 ``training levels'', that introduced participants to the
game mechanisms. In the first training level, players learned the
objective of the game. In the second training level, they discovered how
the character could change its orientation by itself. After this
training phase, participants continued with the main phase of the
experiment.

During the main phase of the experiment, participants played once each
of the four levels (EASY, MEDIUM, HARD or ULTRA), in a random order.
Immediately after the end of a level, participants were given a NASA-TLX
questionnaire to inquire about their mental workload. The questionnaire
took the form of a 9-points Likert scale. As in the original
questionnaire \citep{Hart1988}, it comprised 6 items, that assessed
mental demand, physical demand, temporal demand, performance, effort and
frustration. The experiment lasted approximately 25 minutes and finished
once participants played all 4 levels and filled the corresponding
NASA-TLX questionnaire.

\subsection{Results}\label{results}

\begin{figure}
\centering
\includegraphics[width=0.700\hsize]{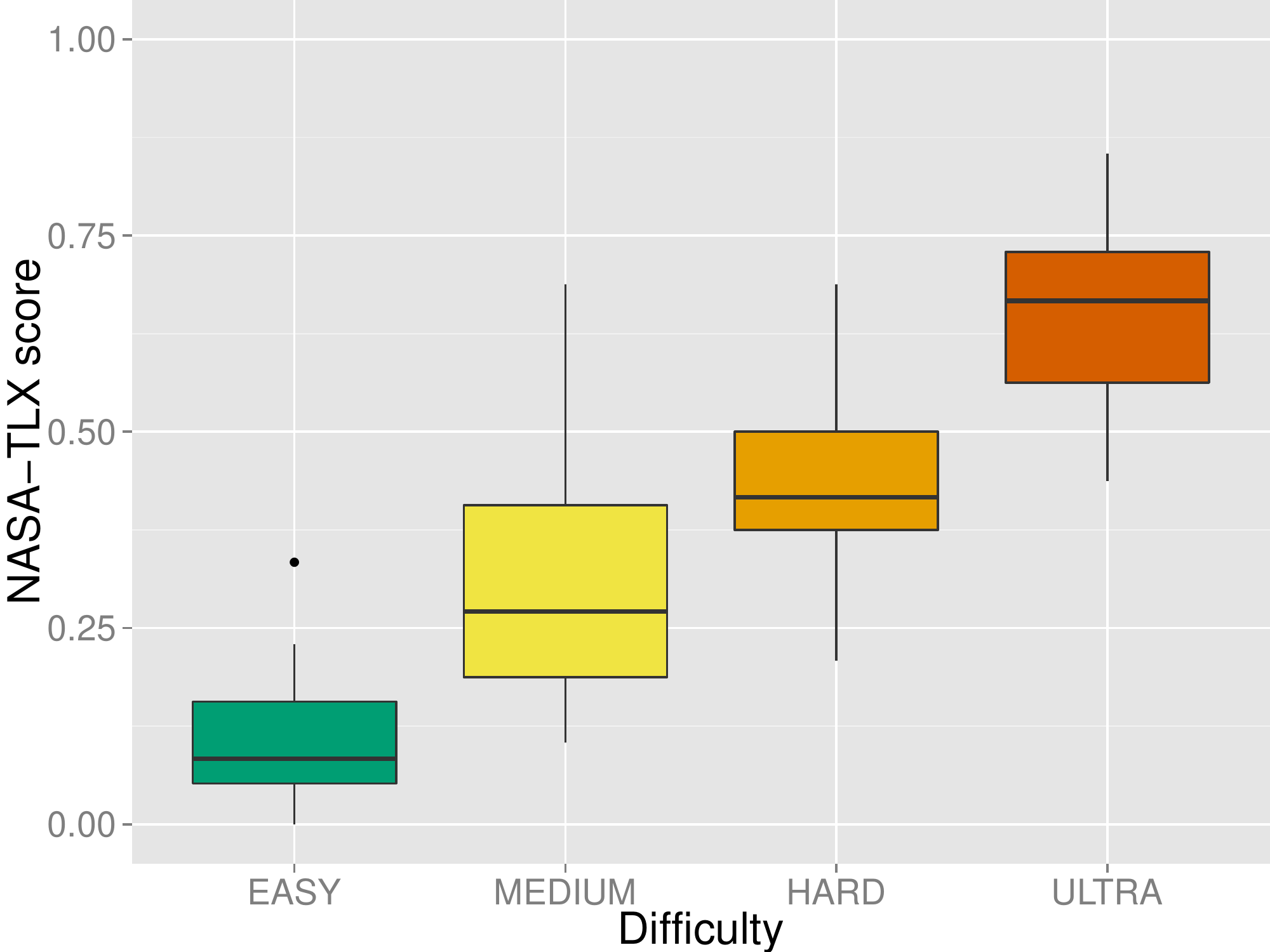}
\caption{NASA-TLX scores obtained during the pilot study. Each
difficulty level differs significantly from the others (p \textless{}
0.01).}\label{fig:nasa-tlx}
\end{figure}

For each participant and each level of difficulty, we averaged the 6
items of the NASA-TLX questionnaire and normalized the scales from
{[}1;9{]} to {[}0;1{]} -- except for the ``performance'' item, that was
normalized from {[}1;9{]} to {[}1;0{]} because its scale is in reverse
order compared to the other items (``1'' for ``good'' and ``9'' for
``poor'').

The resulting averaged scores are: EASY: 0.11 (SD: 0.09); MEDIUM: 0.32
(SD: 0.17); HARD: 0.43 (SD: 0.13); ULTRA: 0.65 (SD: 0.13) -- see
Figure~\ref{fig:nasa-tlx}. A repeated measures analysis of variance
(ANOVA) showed a significant effect of the difficulty factor over the
NASA-TLX scores and a post-hoc pairwise Student's t-test with false
discovery rate (FDR) correction showed that each levels differed
significantly from the others (p \textless{} 0.01).

\subsection{Discussion}\label{discussion}

In this pilot study, we demonstrated through questionnaires that each
difficulty level presented in Table~\ref{tab:difficulties} induces a
different workload level. Hence, we can use our virtual environment as a
baseline to assess the reliability of analogous EEG measures and put
into perspective this new evaluation method.

\section{EEG in practice}\label{eeg-in-practice}

EEG measures the brain activity under the form of electrical currents
\citep{Niedermeyer2005}. To identify mental states from EEG, 3 types of
information can be used:

\begin{itemize}
\itemsep1pt\parskip0pt\parsep0pt
\item
  Frequency domain: oscillations that occur when large groups of neurons
  fire altogether at a similar frequency
\item
  Temporal information: ERPs possess temporal features; positive and
  negative ``peaks'' with varying amplitudes and delays.
\item
  Spatial domain: position of the electrodes that record a specific
  brain activity.
\end{itemize}

However, there is an important variability between people's EEG signals,
and many external factors that could influence EEG recordings
(amplifier's specifications, electrodes exact location, and so on). As
such, it is difficult to identify a universal set of features to
estimate a given mental state, for different sessions and participants.
This is why machine learning is typically used in EEG studies
\citep{Blankertz2010}. With this approach, a calibration phase occurs so
that the system could learn which features are associated to a specific
individual, during a task that is known to induce the studied construct.
Once the calibration is completed, the machine could then use this
knowledge to gain insights about an unknown context, for example a new
interaction technique that one would want to evaluate.

To calibrate workload and attention, we chose to use standard
calibration tasks, validated by the literature, so that our findings
could be easily reproduced. Moreover, as shown later in this paper,
using a single of these tasks to calibrate each construct estimator was
enough to obtain reliable estimations of such constructs during
different and complex interaction tasks.

Concerning attention, we did not develop a dedicated task \emph{per se}
for its calibration. Since the audio probes were already integrated to
our virtual environment, we simply used a specific level of our game.

\subsection{Calibration of workload}\label{calibration-of-workload}

We used the protocol known as the N-back task to induce 2 different
workload levels and calibrate our workload estimator. The N-back task is
a well-known task to induce workload by playing on memory load
\citep{Owen2005}. It showed promising results in \citep{Wobrock2015}
where it could be used to transfer calibration results to a 3D context.

In the N-back task, users watch a sequence of letters on screen, the
letters being displayed one by one. For each letter the user had to
indicate whether the displayed letter was identical or different to the
letter displayed N letters before, using a left or right mouse click
respectively. Hence, users have to remember \emph{n} items at all times.

\begin{figure}
\centering
\includegraphics[width=0.900\hsize]{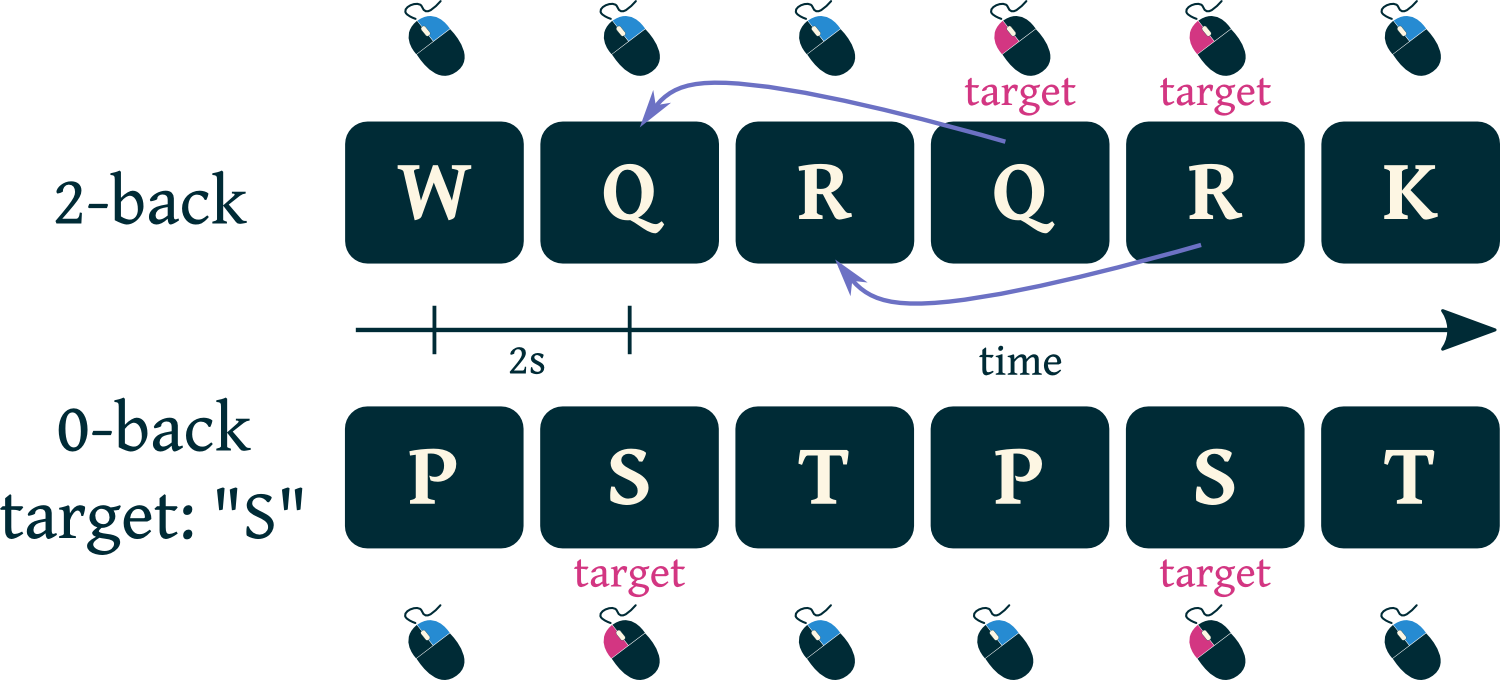}
\caption{Workload calibration task. \emph{Top}: difficult task (2-back
task), the target letter is the one that appeared two steps earlier,
users have to select trials 4 and 5. \emph{Bottom}: easy task (0-back
task), the target letter ``S'' is randomly chosen, users have to select
trials 2 and 5.}\label{fig:n-back}
\end{figure}

We implemented a version similar to \citep{Grimes2008}, removing vowels
to prevent chunking strategies based on phonemes. We used the same time
constraint as in \citep{Wobrock2015}, i.e.~letters appeared for 0.5s,
with an inter-stimulus interval of 1.5s. Each user alternated between
``easy'' blocks with the 0-back task (the user had to identify whether
the current letter was a randomly chosen target letter, e.g. `X') and
``difficult'' blocks with the 2-back task (the user had to identify
whether the current letter was the same letter as the one displayed 2
letters before), see Figure~\ref{fig:n-back}.

Each block contained 60 letters presentations. 4 letters were drawn at
the beginning of a block so that the number of target letters accounted
for 25\% of the trials. Each participant completed 6 blocks, 3 blocks
for each workload level (0-back \emph{vs} 2-back). Therefore, 360
calibration trials (i.e.~one trial being one letter presentation) were
collected for each user, with 180 trials for each workload level
(``low'' \emph{vs} ``high''). This calibration phase takes approximately
12 minutes.

\subsection{Calibration of error
recognition}\label{calibration-of-error-recognition}

\begin{figure}
\centering
\includegraphics[width=0.600\hsize]{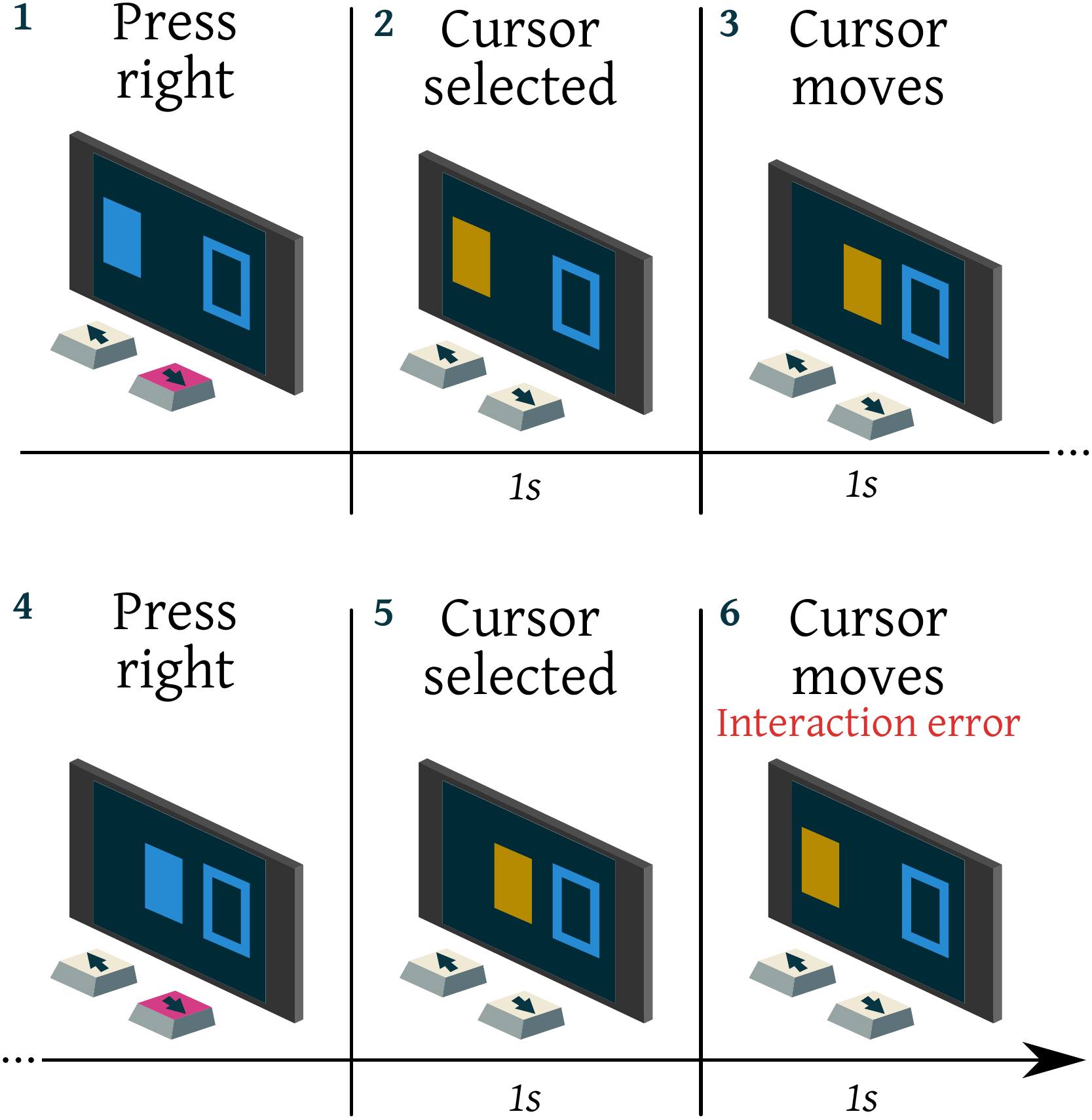}
\caption{Error recognition calibration task. Users control a blue filled
rectangle. They have to move it to an outlined target by pressing the
left or right arrow key. 20\% of the time, the rectangle goes in the
opposite direction, thus causing an interaction
error.}\label{fig:errp-task}
\end{figure}

We replicated the standard protocol described in \citep{Ferrez2008} to
calibrate the system regarding error recognition. The task simulates a
scenario in which users control the movements of a robot. The robot
appears on screen and has to reach a target. At each turn users command
the robot to go right or left in order to reach the target as fast as
possible (with the least steps). However, the robot may understand badly
the given command. This is simulated by some trials during which the
command is (on purpose) erroneously interpreted; hence an interaction
error happens. The ERP that can be seen in EEG following an interaction
error is known as an ``error related potential'', ErrP
\citep{Ferrez2008}. The calibration task is a simplified version of this
scenario: the robot is pictured by a blue rectangle on screen that users
control with the arrow keys, the target is represented by a blue
outline. The robot is constrained to the X axis and along this axis
there are only 7 different positions both for the robot and the target
(see Figure~\ref{fig:errp-task}).

We choose a ratio for the occurrence of interaction errors that is
consistent with the literature. 80\% of the movements matched the actual
key pressed and for the other 20\% the ``robot'' moved in the opposite
direction. It was necessary not to balance both events as too frequent
errors may not be perceived as unexpected anymore, and thus may not lead
to an ErrP. A timer was set to prevent the appearance of artifacts, such
as muscle movements, within EEG recordings (see the Discussion section
for further consideration for artifacts). The rectangle moves 1s after a
key was pressed, and after movement completion users have to wait
another 1s before they could press a key again. The rectangle turned
yellow to tell users they could not control it during that second.

A trial is completed once the robot reaches the target. A trial fails if
after 10 attempts the robot is not yet on target. At the beginning of
each trial, the screen is reinitialized with a random new position for
the robot and the target. The last trial occurred after 350 interactions
were performed. On average this calibration phase lasted 15 minutes.

\subsection{Calibration of attention}\label{calibration-of-attention}

The calibration of attention occurred within a simplified version of the
virtual environment. Users did not have to control the character during
this special level, it was moving by itself through the maze. They were
asked to watch the character and count in their head how many times they
heard the ``odd'' sound, i.e., a high pitched bell lasting 200ms. The
distractor was a low pitched beat of 70ms -- we did not use pure tones
to improve users experience. The pace of the game was adjusted so that a
sound (target or distractor) was played every second. Since the probes
for attention relies to the oddball paradigm, we chose a 20\% likelihood
of appearance for the target event. The calibration lasted about 7
minutes, after 350 sounds were played. Note that participants were
instructed to count the ``odd'' events \emph{only} during the
calibration phase, and \emph{not} during the completion of the 3D maze.

\section{Main study: EEG as an evaluation
method}\label{main-study-eeg-as-an-evaluation-method}

The main study consisted in the evaluation of the game environment with
two different types of interfaces using EEG recordings. As such we
created a 4 (difficulty: EASY, MEDIUM, HARD, ULTRA) $\times$ 2
(interaction: KEYBOARD \emph{vs} TOUCH) within-subject experimental
plan. Our hypotheses are:

\begin{enumerate}
\def\labelenumi{\arabic{enumi}.}
\itemsep1pt\parskip0pt\parsep0pt
\item
  The workload index measured by EEG is higher in TOUCH and increases
  with the difficulty, reflecting NASA-TLX scores obtained during the
  pilot study.
\item
  The attentional resources that participants assign to the sounds
  decrease as the difficulty increases.
\item
  The TOUCH condition induces a higher number of interaction errors
  compared to the KEYBOARD condition.
\end{enumerate}

The gaming phase was split into two sequences, one for each interaction
technique. To avoid a too tedious experiment, participants alternated
between game sessions and the 3 calibration tasks (workload, attention
and error recognition). Since the analysis were performed offline, there
was no need to cluster all the calibrations at the beginning of the
experiment.

\begin{figure}
\centering
\includegraphics[]{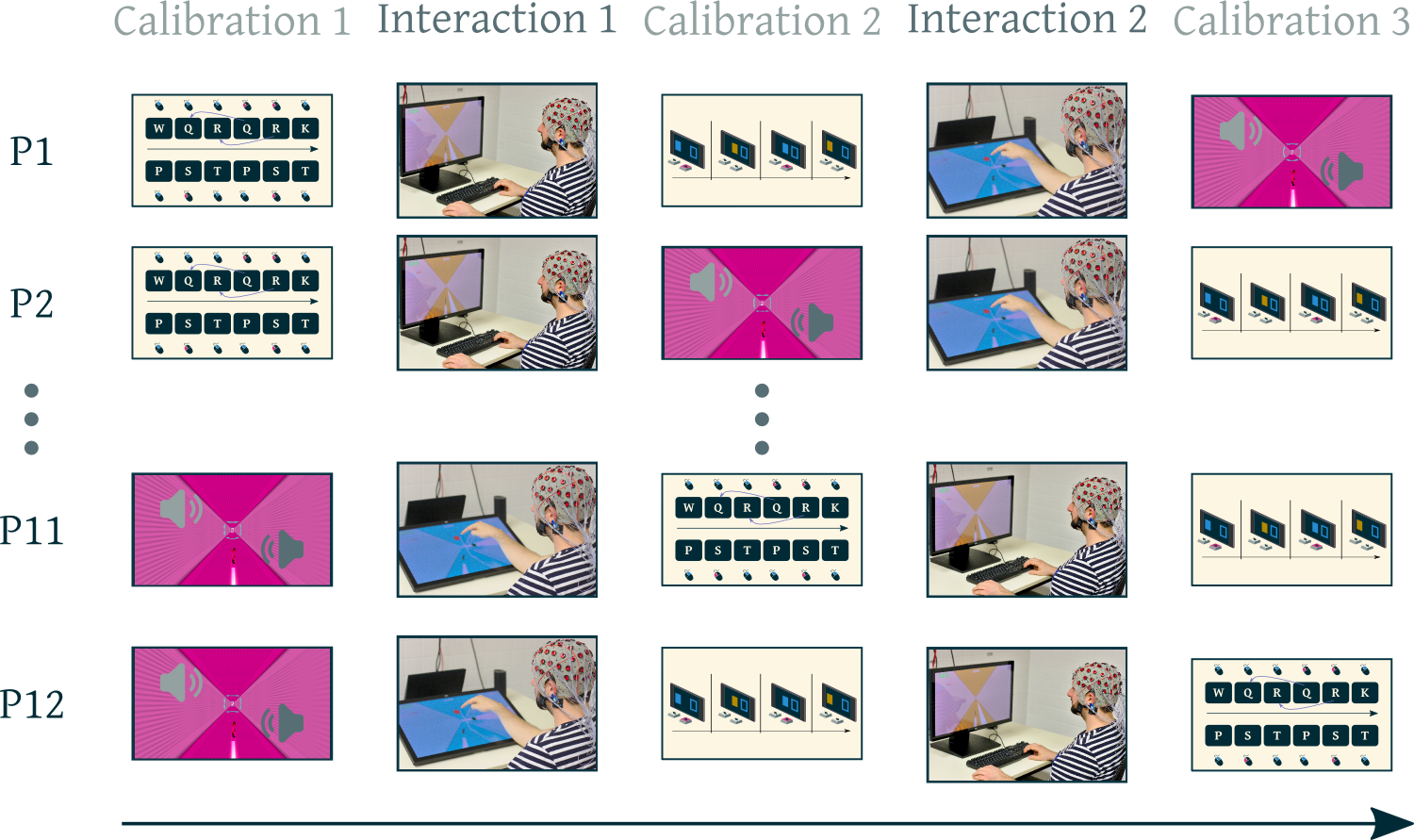}
\caption{The order of the 3 calibration tasks and 2 interaction
techniques was counter-balanced between the 12 participants to improve
engagement.}\label{fig:latin}
\end{figure}

The order of the gaming sessions and calibration phases was
counter-balanced between participants following a latin square (see
Figure~\ref{fig:latin}). After the experiment, the signals gathered from
the calibration tasks were processed in order to evaluate both the
virtual environment (difficulty levels) and the chosen interaction
techniques.

\subsection{Apparatus}\label{apparatus}

EEG signals were acquired at 512Hz with 2 g.tec g.USBamp amplifiers. We
used 32 electrodes placed at the AF3, AFz, AF4, F7, F3, Fz, F4, F8, FC3,
FCz, FC4, C5, C3, C1, Cz, C2, C4, C6, CP3, CPz, CP4, P7, P3, Pz, P4, P8,
PO7, POz, PO8, O1, Oz and O2 sites.

12 participants took part in this study (3 females), mean age 26.25 (SD:
3.70). All of them reported a daily use of tactile interfaces. The
experiment occurred in a quiet environment, isolated from the outside.
There were two experimenters in the room and the procedure comprised the
following steps:

\begin{enumerate}
\def\labelenumi{\arabic{enumi}.}
\itemsep1pt\parskip0pt\parsep0pt
\item
  Participants entered the room, read and signed an informed consent
  form and filled a demographic questionnaire.
\item
  While one of the experimenter installed an EEG cap onto participants'
  heads, the other experimenter introduced participants to the virtual
  environment. They played 2 training levels and the 4 main levels in an
  increasing order of difficulty. They could redo some levels if they
  did not feel confident enough.
\item
  One of the 3 calibration tasks occurred (workload, attention or error
  recognition).
\item
  Participants played the game using one of the 2 interaction techniques
  (KEYBOARD or TOUCH). The four levels of difficulty (EASY, MEDIUM,
  HARD, ULTRA) appeared twice during the session, in a random order. For
  TOUCH, a dedicated training session occurred beforehand so that
  participant could get used to this interaction technique.
\item
  Another calibration task occurred, different from step 3.
\item
  Participants tested the second interaction technique. As in step 4,
  TOUCH was preceded by a training session, that lasted until
  participants felt confident enough to proceed to the main task.
\item
  Participants performed the last remaining calibration task.
\end{enumerate}

A game session (steps 4 and 6) took approximately 20 minutes and the
whole experiment lasted 2 hours.

\subsection{EEG Analyses}\label{eeg-analyses}

The calibration tasks were used to train a classifier specific to each
of the studied construct. Classifiers were calibrated separately for
each participant which ensured maximal EEG classification performances.
We used EEGLAB 13.4.4b\footnote{http://sccn.ucsd.edu/eeglab/} and Matlab
R2014a to process EEG signals offline. EEG features associated to
workload relate to the frequency domain while the features associated to
attention and error recognition relate to temporal information, as
detailed below.

\subsubsection{Processing workload}\label{processing-workload}

From the signals collected during the N-back tasks, we extracted EEG
features from each 2s time window following a letter presentation. We
used each of these time windows as an example to calibrate our
classifier, whose objective was to learn whether these features
represented a low workload level (induced by the 0-back task) or a high
workload level (induced by the 2-back task). Once calibrated, this
classifier can be used to estimate workload levels on new data, here
while our users were interacting with the virtual environment.

As in \citep{Wobrock2015}, we filtered EEG signals in the delta (1-3
Hz), theta (4-6 Hz), alpha (7-13 Hz), beta (14-25 Hz) and gamma (26-40
Hz) bands. To reduce features dimensionality, we used for each band a
set of Common Spatial Patterns (CSP) spatial filters. That way, we
reduced the 32 original channels down to 6 ``virtual'' channels that
maximize the differences between the two workload levels
\citep{Ramoser2000}. Since the calibration (N-back task) and use
contexts (virtual environment) differs substantially, we used a
regularized version of these filters called stationary subspace CSP
(SSCSP) \citep{Wobrock2015}. SSCSP filters are more robust to changes
between contexts since they take into account the distributions of the
EEG signals recorded during both the calibration and the use contexts
(in an unsupervised way, i.e.~without considering the expected workload
levels) to estimate spatial filters whose resulting signals are stable
across contexts (see \citep{Wobrock2015} for details). Finally, for each
frequency band and spatial filter, we used the average band power of the
filtered EEG signals as feature. This resulted in 30 EEG features (5
bands $\times$ 6 spatial filters per band).

\subsubsection{Processing attention and error
recognition}\label{processing-attention-and-error-recognition}

Since both attention and error recognition can be measured in ERPs, they
share the same signal processing. We selected time windows of 1s,
starting at the event of interest (i.e.~sounds for attention,
rectangle's movements for error recognition). In order to utilize
temporal information, feature extraction relied on regularized Eigen
Fisher spatial filters (REFSF) method \citep{Hoffmann2006}. Thanks to
this spatial filter, specifically designed for ERPs classification, the
32 EEG channels were reduced to a set of 5 channels. We then decimated
the signal by a factor 32. The ``decimate'' function of Matlab, that
applies a low-pass filter before decimation to prevent aliasing, was
used. As a result, there was 80 features by epoch (5 channels $\times$
512Hz $\times$ 1s / 32).

\subsubsection{Classification}\label{classification}

We used a shrinkage LDA (linear discriminant analysis) as a classifier
since it is more efficient than the regular LDA with a high number of
features \citep{Ledoit2004}.

For each construct there was two steps: first we used the data collected
during the calibration tasks to estimate the performance of the
classifiers. Second, we studied the output of the different classifiers
to evaluate the virtual environment.

To assess the classifiers' performance on the calibration data, we used
4-fold cross-validation (CV). More precisely, we split the collected
data into 4 parts of equal size, selecting trials randomly, used 3 parts
to calibrate the classifiers and tested the resulting classifiers on the
unseen data from the remaining part. This process occurred 3 more times
so that in the end, each part was used once as test data. Finally, we
averaged the obtained classification performances. The performance was
measured using the area under the receiver-operating characteristic
curve (AUROCC). The AUROCC is a metric that is robust against unbalanced
classes, as it is the case with attention and error recognition (20\% of
targets, 80\% of distractors). A score of ``1'' means a perfect
classification, a score of ``0.5'' is chance.

Once the classifiers were trained thanks to the calibrations tasks, we
could use them on the EEG signals acquired while participants were
interacting with the virtual environment, to estimate the different
constructs values.

For workload, we used 2s long sliding time windows that were overlapping
by 1s, to extract signals and feed the classifier. From the outputs that
was produced by the LDA classifier for each participant (i.e., the
distance to the separating hyperplane), we first removed outliers by
iteratively removing one outlier at a time using a Grubb's test with
$p=0.05$, until no more outlier was detected \citep{Grubbs1969}. We then
normalized the outlier-free scores between -1 and +1. As such, for all
participants a workload index close to +1 represents the highest mental
workload they had to endure while they were playing. It should come
close to the 2-back condition of the calibration phase. On the opposite,
a workload index close to -1 denotes the lowest workload, similar to the
0-back condition.

The process was similar for attention, but we only extracted epochs that
corresponded to the target stimuli onset, i.e.~when the high pitch sound
was played. Note that contrary to \citep{Burns2015}, that studied the
amplitudes of ERPs and did not use the data gathered during the
calibration phase, here we kept the machine learning approach. As such,
the resulting scores can be seen as a confidence index of the LDA
classifier about whether participants noticed odd events while they were
playing.

As for the classifier dedicated to error recognition, the processing
differs. Indeed, we could not assume which interaction yielded or not an
interaction error, i.e.~if and when participants perceived a discrepancy
between what they intended to do and what occurred. Consequently, we
simply counted over an entire game session the number of times the
classifier labelled an interaction as being erroneous in the eye of the
participants.

\subsection{Results}\label{results-1}

Unless otherwise noted, we tested for significance using repeated
measures ANOVA. For significant main effects, we used post-hoc pairwise
Student's t-test with FDR correction.

\begin{table*}
\centering
\begin{tabular}{llllllllllllll}
\toprule\addlinespace
Construct & P1 & P2 & P3 & P4 & P5 & P6 & P7 & P8 & P9 & P10 & P11 & P12
& Average\tabularnewline
\midrule
Workload & 0.85 & 0.93 & 0.98 & 0.95 & 0.97 & 0.97 & 0.79 & 0.87 & 0.87
& 0.98 & 0.95 & 0.94 & 0.92\tabularnewline
Attention & 0.83 & 0.82 & 0.96 & 0.81 & 0.85 & 0.90 & 0.82 & 0.82 & 0.86
& 0.92 & 0.88 & 0.83 & 0.86\tabularnewline
Error recognition & 0.88 & 0.57 & 0.90 & 0.90 & 0.86 & 0.90 & 0.78 &
0.80 & 0.88 & 0.78 & 0.85 & 0.74 & 0.82\tabularnewline
\bottomrule
\end{tabular}
\caption{Classification accuracy during the calibration tasks for the 3
measured constructs (AUROCC scores).}\label{tab:classification}
\end{table*}

\begin{figure}
\centering
\subfloat[\label{fig:workload-stats}]{\includegraphics[width=1.000\hsize]{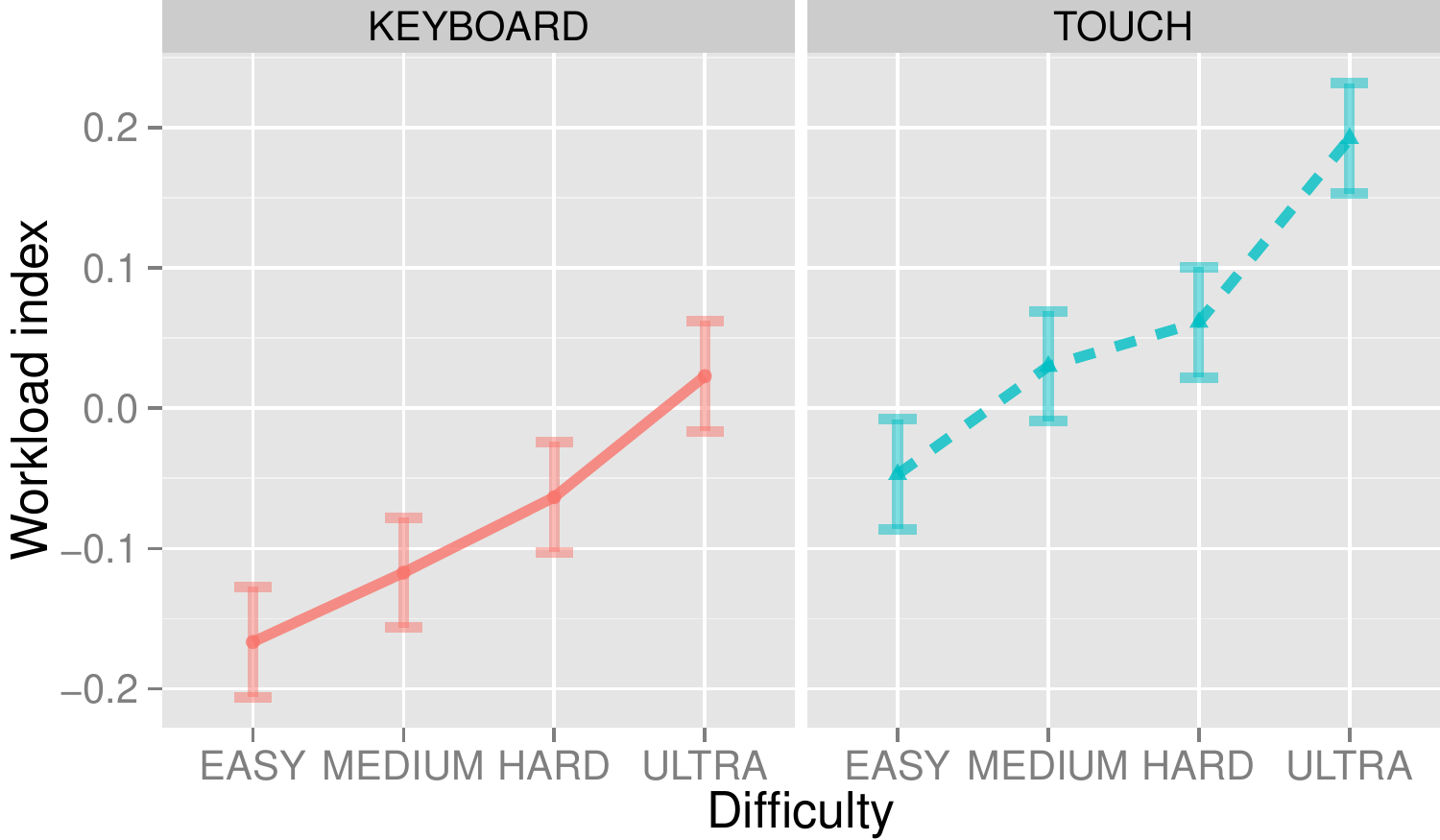}}
\\
\subfloat[\label{fig:attention-stats}]{\includegraphics[width=0.660\hsize]{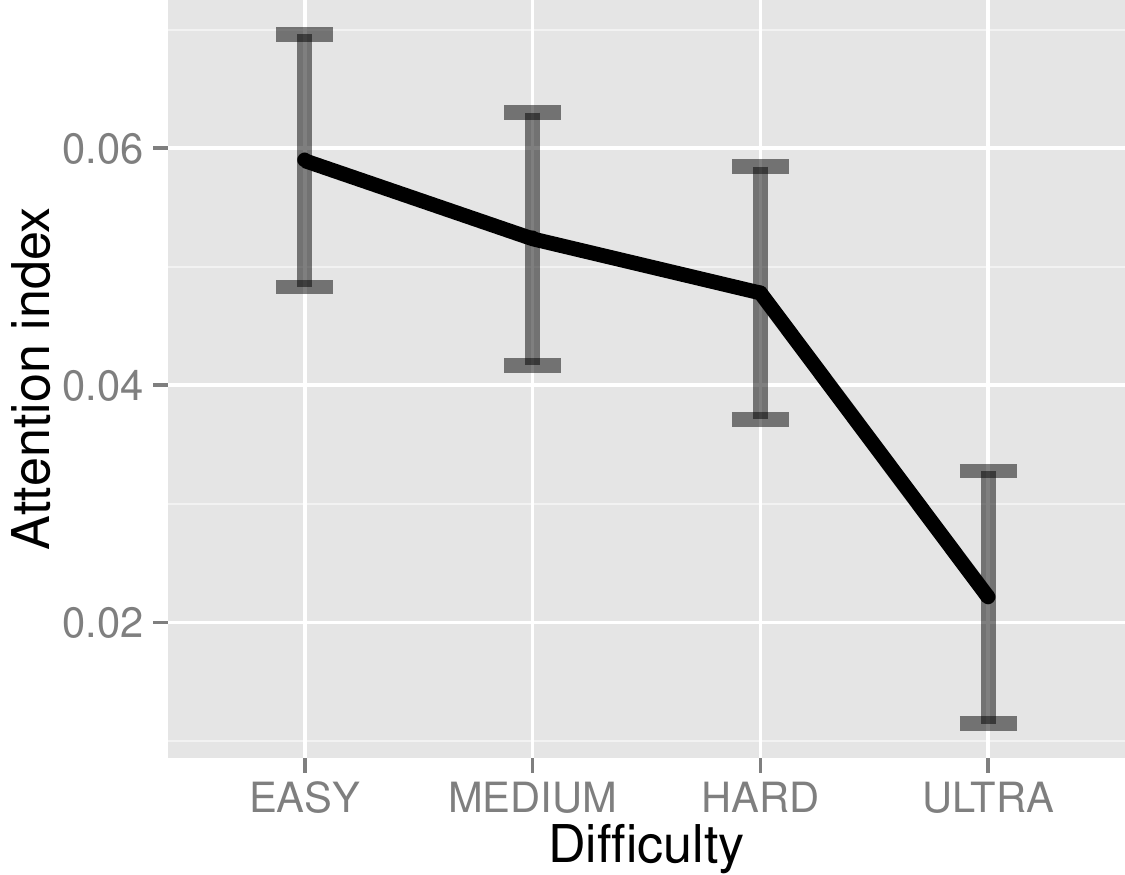}}
\subfloat[\label{fig:errp-stats}]{\includegraphics[width=0.330\hsize]{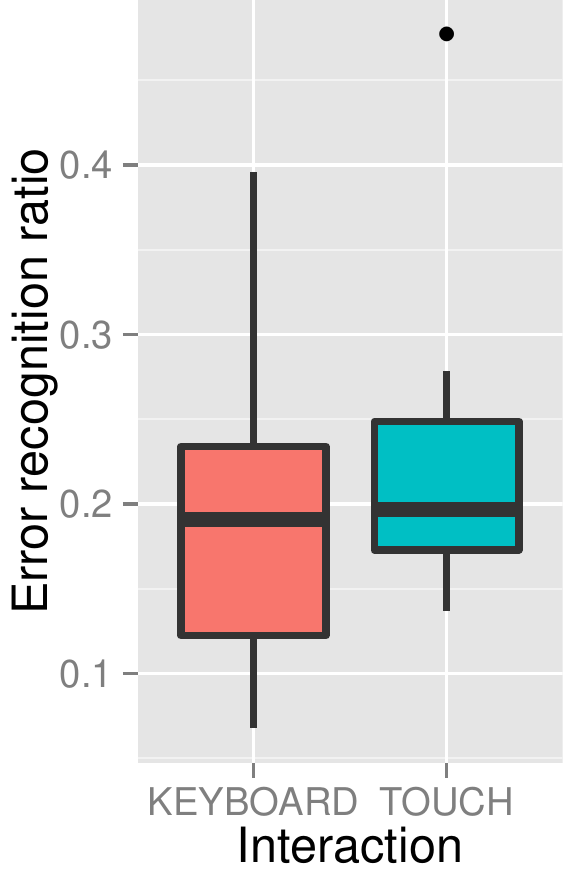}}
\caption{EEG measures. \emph{a}: The workload index significantly
differs across difficulties and between interaction techniques.
\emph{b}: The attention index significantly differs across difficulties.
\emph{c}: The number of interaction errors differs by a tendency between
KEYBOARD and TOUCH.}\label{fig:EEG-stats}
\end{figure}

\subsubsection{Workload}\label{workload}

On average, the classifier AUROCC score during the training task was
0.92 (SD: 0.06) -- see Table~\ref{tab:classification}. Over the test set
there were on average 2171 data points per subject across all condition
(time windows). The statistical analysis of the classifier output during
the game session showed a significant effect of the difficulty factor (p
\textless{} 0.01); the workload index increasing along the difficulty of
the levels (Figure~\ref{fig:workload-stats}). The post-hoc analysis
showed that all difficulty levels significantly differs one from the
other with p \textless{} 0.01; except for the MEDIUM level, which
differs from EASY with p \textless{} 0.05 and with HARD only by a margin
(p = 0.11). There was a significant effect of the interaction factor as
well (p \textless{} 0.01), the workload being higher on average during
the TOUCH condition. There was no interaction between difficulty and
interaction factors.

\subsubsection{Attention}\label{attention}

On average, the classifier AUROCC score during the training task was
0.86 (SD: 0.05) -- see Table~\ref{tab:classification}. Over the test set
there were on average 497 data points per subject across all conditions
(odd events). The statistical analysis of the classifier output during
the game session showed a significant effect of the difficulty factor (p
\textless{} 0.01) but not of the interaction factor. The attention index
decreases as the difficulty increases
(Figure~\ref{fig:attention-stats}). The post-hoc analysis showed that
the ULTRA level significantly differs from the others (p \textless{}
0.05).

\subsubsection{Error recognition}\label{error-recognition}

On average, the classifier AUROCC score during the training task was
0.82 (SD: 0.10) -- see Table~\ref{tab:classification}. Over the test set
there were on average 388 data points per subject across all conditions
(interactions). Due to the nature of the data (numbers of interaction
errors across entire game sessions), we used a one-tailed Wilcoxon
Signed Rank Test to stress our hypothesis. The number of interaction
errors differs by a tendency (p = 0.08) between the KEYBOARD and the
TOUCH conditions. 19\% of the interactions (SD: 9\%) were labelled as
interaction errors by the classifier for KEYBOARD \emph{vs} 22\% (SD:
9\%) for TOUCH (Figure~\ref{fig:errp-stats}).

\subsection{Behavioral measures}\label{behavioral-measures}

Besides EEG metrics, we had the opportunity to study participants'
reaction time and performance so as to get a clearer picture of their
user experience.

\begin{figure}
\centering
\subfloat[\label{fig:time-stats}]{\includegraphics[width=0.480\hsize]{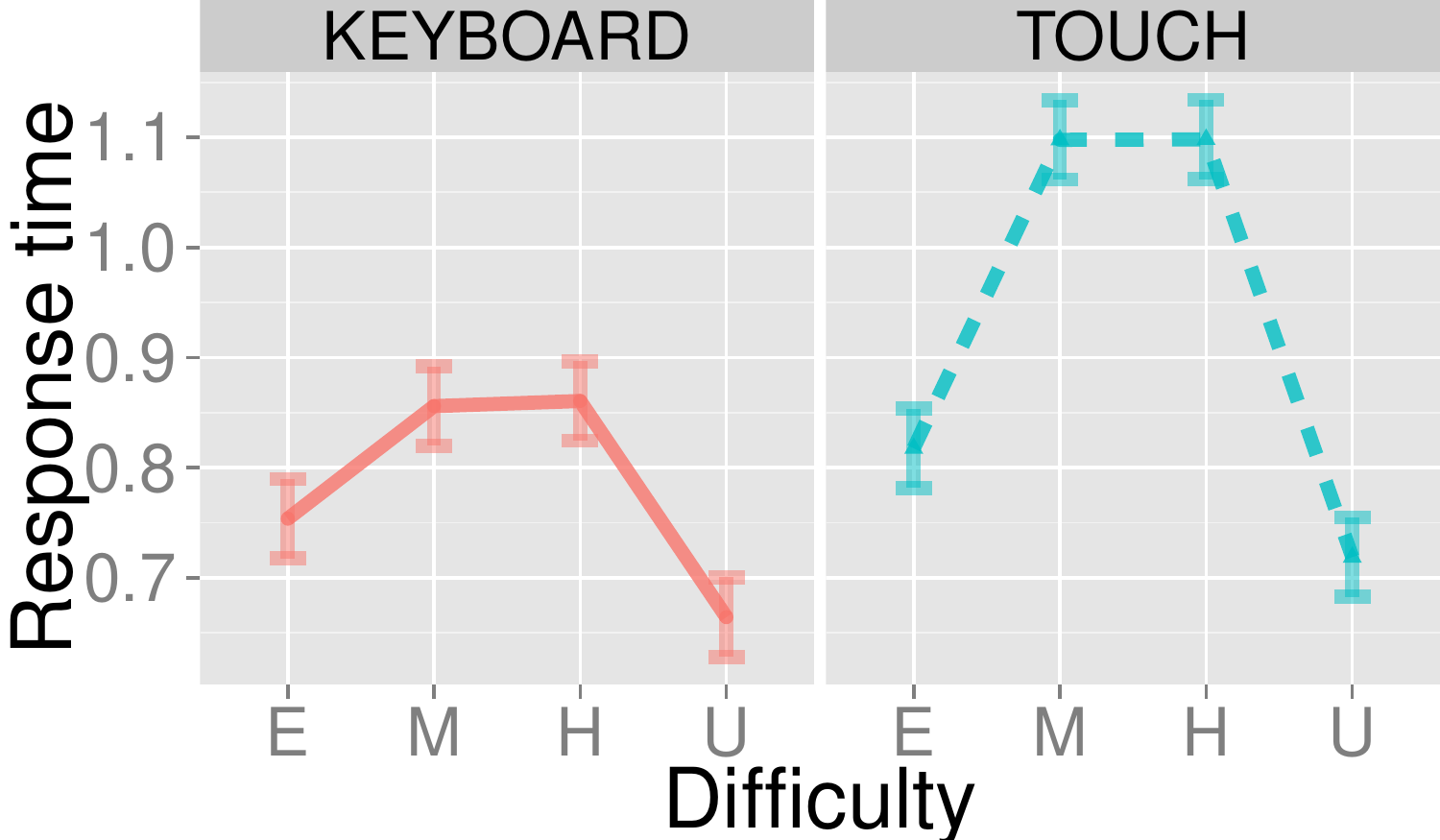}}
\subfloat[\label{fig:perf-stats}]{\includegraphics[width=0.480\hsize]{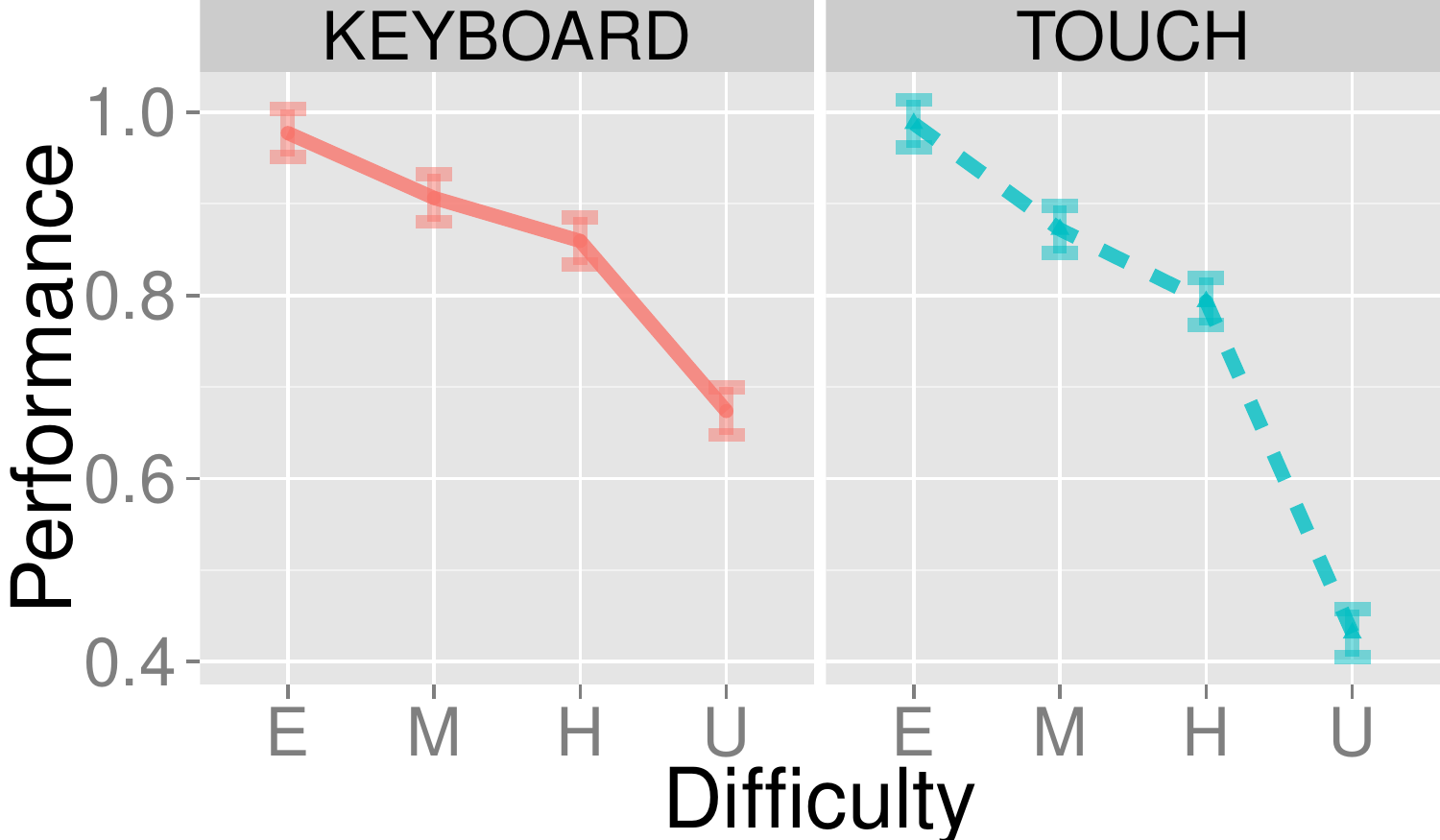}}
\caption{Behavioral measures: reaction time in seconds (left) and
performance (proportion of correctly selected directions -- right)
significantly differs between difficulty levels and interactions. E:
EASY, M: MEDIUM, H: HARD, U: ULTRA.}\label{fig:behav-stats}
\end{figure}

\subsubsection{Reaction time}\label{reaction-time}

There was a significant effect of both the difficulty and interaction
factors, as well as an interaction effect between them (p \textless{}
0.01). Post-hoc tests showed that all difficulty levels differ from one
another (p \textless{} 0.01), except for MEDIUM and HARD, which do not
differ significantly (p = 0.91). The mean reaction times were
respectively for EASY, MEDIUM, HARD and ULTRA: 0.78s (SD: 0.14), 0.97s
(0.18), 0.98s (0.15) and 0.69s (0.06). The mean reaction time was 0.78
(0.12) for KEYBOARD and 0.93 (0.13) for TOUCH. See
Figure~\ref{fig:time-stats}. Note that users had less time to respond
during higher difficulty levels.

\subsubsection{Performance}\label{performance}

The performance was computed as the ratio between the number of correct
selections and the total number of interactions. There was a significant
effect of both the difficulty and interaction factors, as well as an
interaction effect between them (p \textless{} 0.01). Post-hoc tests
showed that all difficulty levels differ from one another (p \textless{}
0.01). The mean performance was respectively for EASY, MEDIUM, HARD and
ULTRA: 98\% (SD: 3), 89\% (12), 83\% (17) and 55\% (21). The Mean
performance was 85\% (13) for KEYBOARD and 77\% (13) for TOUCH. See
Figure~\ref{fig:perf-stats}.

\subsection{Discussion}\label{discussion-1}

Most of the main hypotheses are verified. The workload index as computed
with EEG showed significant differences that match the intended design
of the difficulty levels. It was also shown that in the highest
difficulty the attention level of participants toward external stimuli
was significantly lower -- i.e.~inattentional blindness increased.
Concerning the interaction techniques, the number of interaction errors
as measured by EEG was higher with the TOUCH condition, but this is a
tendency and not a significant effect. The workload index, on the other
hand, was significantly higher in the TOUCH condition compared to the
KEYBOARD condition.

Thanks to the ground truth obtained during the pilot study with the
NASA-TLX questionnaire, these results validate the use of a workload
index measured by EEG for HCI evaluation and set the path for two other
constructs: attention and error recognition. Beside the evaluation of
the content (i.e.~difficulty levels) we were able to compare two
interaction techniques. These are promising results for those who seek
to assess how intuitive a UI is with exocentric measures
\citep{Frey2014a}.

In this study, we chose to use the particularity of the touch screen to
make the task \emph{more} difficult. Indeed, while we used a touch
screen for its possibility of direct manipulation, we kept the character
as a frame of reference, resulting in input commands that were
(patently) not co-localized with output directions. Besides results
denoting the differences between the conditions, participants also
spontaneously reported how non intuitive this condition was. We wanted
to investigate our evaluation method on a salient difference at first.
Then our framework could well be employed to go further; for example
seeking physiological differences between direct and indirect
manipulation interfaces in more traditional tasks.

It is interesting to note how those EEG measures could be combined with
existing methods to broaden the overall comprehension of the user
experience. For instance, while we did show significant differences
across difficulty levels and between interaction techniques with
behavioral measures (reaction time and performance index), EEG measures
could help to understand the underlying mechanisms. Because we have a
more direct access to brain activity, we can make assumptions about the
cause of observed behaviors. For example participants' worse performance
with TOUCH than with KEYBOARD could be due to the fact that they
anticipate less the outcomes of their actions (more interaction errors);
the higher reaction time may not only be caused by the interaction
technique \emph{per se}, but by a higher workload. And while
participants manage to cope with the fast pace of the ULTRA level (the
smallest reaction times), the increase in perceptual load lower their
awareness to task-irrelevant stimuli.

Additionally, we can observe that the performances obtained at the EASY,
MEDIUM and HARD levels are very similar with the keyboard and the touch
screen (see Figure~\ref{fig:perf-stats}). However, EEG analyses revealed
that the workload was significantly higher in the TOUCH condition,
meaning that users had to allocate significantly more cognitive
resources to reach the same performance. This further highlights that
EEG-based measures do bring additional information that can complement
traditional evaluations such a behavioral measures.

Measuring users' cognitive processes such as workload and attention may
prove particularly useful to assess 3D user interfaces (3DUI), since
they are known to be more cognitively demanding. They require users to
perform 3D mental rotation tasks to successfully manipulate objects or
to orientate themselves in the 3D environment. Moreover, the usual need
for a mapping between the user inputs (with limited degrees-of-freedom
-- e.g., only 2 for a mouse) and the corresponding actions on 3D objects
(with typically 6 degrees-of-freedom), makes 3DUI usually difficult to
assess and design. We reproduced part of this problematic with our game
environment and obtained coherent results from EEG measures.

\begin{figure}
\centering
\includegraphics[]{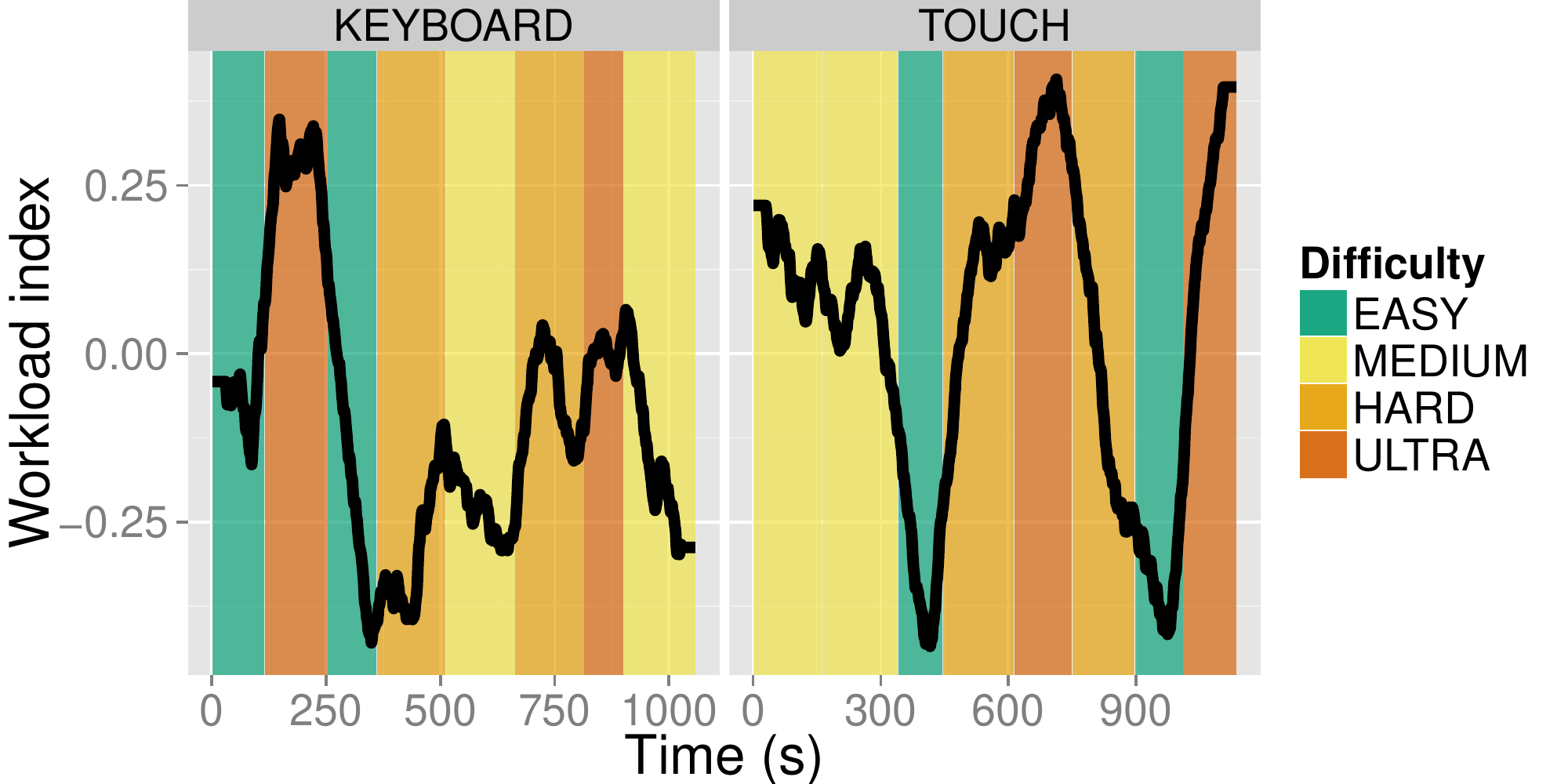}
\caption{Workload index over time for participant 3 -- 60s smoothing
window. \emph{Left}: KEYBOARD condition, \emph{right}: TOUCH condition.
Background color represents the corresponding difficulty
level.}\label{fig:workload-continuous}
\end{figure}

Above all, an evaluation method based on EEG enables a continuous
monitoring of users. The intended use case of our framework is to enroll
dedicated testers that would wear the EEG equipment and perform well
during the calibration tasks. As a matter of fact, the best performer
during workload calibration (participant 3 in
Table~\ref{tab:classification}) shows patterns that clearly meet the
expectations concerning both difficulty levels and interactions, as
pictured in Figure~\ref{fig:workload-continuous}.

\section{Limitations and future
challenges}\label{limitations-and-future-challenges}

Although using EEG measures as an evaluation method for HCI was proven
conclusive regarding workload -- we obtained a continuous index on par
with a ground truth based on traditional questionnaires -- the two other
constructs we studied could benefit from further improvements.

Despite the direct interaction (TOUCH) being more disorienting for users
than the indirect one (KEYBOARD), the recognition of interaction errors
differed only by a tendency. This could be explained by the fact that
the calibration task was too dissimilar to the virtual environment.
Notably, while there was few and slow paced events during the
calibration, users were confronted to many stimuli while they were
playing, hence overlapping ERPs must have appeared within EEG, which may
have disturbed the classifier. A calibration task closer to real-life
scenarios than the one described in \citep{Ferrez2008} should be
envisioned. Such task should remain generic in order to facilitate the
dissemination of EEG as an evaluation method for HCI.

Signal processing could also facilitate the transfer of the
classification between a standard task and the evaluated HCI. Indeed, if
our results demonstrate that the EEG classification of workload could be
transferred from the N-back tasks to a dissimilar virtual environment
and user interface, we benefited from spatial filters that specifically
take into account the variance between calibration contexts and use
contexts -- the stationary subspace CSP \citep{Wobrock2015}. Since ERPs
may also slightly differ in amplitudes and delays between calibration
and use contexts, in the future, it would be worth designing similar
approaches to optimize temporal or spatial filters for ERPs as well.

The reliability of mental states measures is strongly correlated to the
quality of EEG signals. Interestingly enough, participants' mindset
during the recordings is one of the factors influencing EEG signals.
Their awareness and involvement toward the tasks improve system
accuracy. The form of the calibration tasks could be enhanced to engage
more users, for example through gamification \citep{Flatla2011} -- and
our virtual environment proved to be suitable to do so. Whereas our
participants were volunteers enrolled among students, in the end the
outcome of an evaluation method based on EEG should be strengthened by
recruiting dedicated testers, using as selection criteria how reliably
the different constructs could be estimated from their EEG signals
during calibration tasks.

Finally, one should acknowledge that when it comes to recordings as
sensitive as EEG, artifacts such as the ones induced by muscular
activity are of major concern. The way we prevented the appearance of
such bias in the present study is threefold. 1) The hardware we used --
active electrodes with Ag/AgCl coating -- is robust to cable movements,
see e.g., \citep{Wilson2012a}. 2) The classifiers were trained on
features not related to motion artifacts or motor cortex activation. 3)
The position of the screen during the ``touch'' condition minimized
participants' motion, and gestures occurred mostly before the time
window used for detecting interaction errors. These precautions are
important for the technology to be correctly apprehended.

To further control for any bias in our protocol, we ran a batch of
simulations where the labels of the calibration tasks had been randomly
shuffled, similarly to the verification process described in
\citep{Wobrock2015}. Should artifacts bias our classifiers, differences
would have appeared between the KEYBOARD and TOUCH conditions even with
such random training. Among the 20 simulations that ran for each of the
3 constructs (workload, attention, error recognition), none yielded
significant differences.

\section{Conclusion}\label{conclusion}

In this paper, we demonstrated how brain signals -- recorded by means of
electroencephalography -- could be put into practice in order to obtain
a continuous evaluation of different interaction techniques, for
assessing their ergonomic pros and cons. In particular, we validated an
EEG-based workload estimator that does not necessitate to modify the
existing software. Furthermore, we showed how users' attention level
could be evaluated using background stimuli, such as sounds. Finally, we
investigated how the recognition of interaction errors could help to
determine the best user interface.

Being able to estimate these three constructs -- workload, attention and
error recognition -- continuously during realistic and complex
interaction tasks opened new possibilities. Notably, it enabled us to
obtain additional and more objective metrics of user experience, based
on the users' cognitive processes. It also provided us with additional
insights that traditional measures (e.g., behavioral measures) could not
reveal. To sum up, this suggests that combined with existing evaluation
methods, EEG-based evaluation tools such as the ones proposed here can
help to understand better the overall user experience. We hope that the
increasing availability of EEG devices will foster such approaches and
benefit the HCI field.

\balance{}

\bibliography{biblio}

\end{document}